\documentclass[english,twocolumn, preprintnumbers, aps, showpacs, prd]{revtex4}
\usepackage[T1]{fontenc}
\usepackage[latin1]{inputenc}

\makeatletter

\newcommand{\half}{\mbox{$\textstyle \frac{1}{2}$}}

\newcommand{\myfrac}[2]{\mbox{$\textstyle \frac{#1}{#2}$}}

\usepackage{babel}
\makeatother
\begin{document}

\preprint{LMU-ASC 34/06}

\title{A rule of thumb for cosmological backreaction}

\author{M. Parry}

\affiliation{Arnold Sommerfeld Center, Department für Physik, Ludwig-Maximilians-Universität
München, Theresienstr. 37, D-80333 München, Germany}

\date{10 May 2006}

\begin{abstract}
In the context of second order perturbation theory, cosmological backreaction
is seen to rescale both time and the scale factor. The issue of the
homogeneous limit of long-wavelength perturbations is addressed and
backreaction is quantified in terms of a gauge-invariant metric function
that is the true physical degree of freedom in the homogeneous limit.
The time integral of this metric function controls whether backreaction
hastens or delays the expansion of the universe. As an example, late-time
acceleration of the universe is shown to be inconsistent with a perturbative
approach. Any tendency to accelerate the expansion requires negative
non-adiabatic pressure fluctuations.
\end{abstract}


\maketitle

\section{Introduction}

The application of cosmological perturbation theory to primordial
fluctuations and the cosmic microwave background has been outstandingly
successful \cite{Spergel:2006hy}. Since inhomogeneities tend to grow
in the universe, however, it is clear that first order perturbation
theory will not remain valid. The question then arises: to what extent
do these inhomogeneities act back on the expansion of the universe?
Aligned with this question is the expectation that as observations
become increasingly precise one might see second order effects at
early times \cite{Acquaviva:2002ud,Bartolo:2004if,Martineau:2005aa,Bartolo:2005fp,Finelli:2006wk}.
Consequently, the issue of cosmology perturbations to second order
and beyond has been the subject of considerable inquiry%
\footnote{Space does not allow us to mention the work on fully nonlinear cosmological
perturbations. The interested reader may turn to \cite{Easther:1999ws,Afshordi:2000nr,Langlois:2005qp}
and references therein.%
}.

The treatment of cosmological perturbations to second order was first
carried out in the context of backreaction \cite{Futamase:1988aa,Futamase:1989aa}.
This initiated the general study of inhomogeneities and averaging
in cosmology \cite{Bildhauer:1991aa,Zalaletdinov:1992cg,Buchert:1995fz,Russ:1996km,Zalaletdinov:1996aj},
and culminated in a beautiful series of papers \cite{Buchert:1999er,Buchert:2001sa,Buchert:2002ht,Buchert:2002ij}
(see also \cite{Ellis:2005uz}) highlighting the fact that there is
both a kinematical backreaction because spatial averaging and time
evolution do not commute, and a curvature backreaction because spatial
averaging and spatial rescaling do not commute (see also \cite{Mars:1997jy}).
At the very least, therefore, one should expect a rescaling of the
cosmological parameters; in the extreme limit, the evolution of the
universe can be mischaracterized by imposing on it the FLRW framework.

One of the deficiencies of the early work on second order perturbations
was that it was not gauge-invariant and, hence, was difficult to interpret
reliably. After the mathematical foundation of gauge-invariant perturbation
theory to all orders was established \cite{Mukhanov:1996ak,Bruni:1996im,Abramo:1997hu},
work began in earnest on quantifying backreaction and on identifying
conservation laws for long-wavelength fluctuations \cite{Rigopoulos:2002mc,Lyth:2003im,Noh:2004bc,Malik:2003mv,Vernizzi:2004nc,Malik:2005cy}.
It was shown \cite{Abramo:1997hu} that backreaction of long-wavelength
modes behaves like a negative cosmological constant and it was anticipated
that there could be a dynamical relaxation mechanism for the cosmological
constant \cite{Brandenberger:1999su,Brandenberger:2002sk}. Subsequently,
there was some debate as to whether such backreaction was physically
measurable \cite{Unruh:1998ic,Abramo:2001dc,Afshordi:2000nr,Geshnizjani:2002wp}
and the analysis of backreaction was tightened. Specifically, it was
noted \cite{Geshnizjani:2003cn} that one requires a physical clock,
e.g. a secondary scalar field, to which a physical observable can
be referred.

The most recent direction taken by the study of backreaction concerns
the late-time acceleration of the universe \cite{Riess:1998cb,Perlmutter:1998np}.
It suffices to say that arguments are forcefully put (e.g. \cite{Kolb:2004am,Kolb:2004jg,Flanagan:2005dk,Geshnizjani:2005ce,Ishibashi:2005sj,Buchert:2005kj,Rasanen:2005zy,Siegel:2005xu,Martineau:2005zu})
both for and against the idea that backreaction can be responsible
for late-time acceleration.

This paper is a conservative attempt to apply two lessons from the
study of backreaction. The first is that the scale factor and other
cosmological parameters deduced from observations are rescaled quantities
and not the ``bare'' quantities entering the perturbative formalism.
The second lesson is that if extremely long-wavelength modes are the
primary contribution to backreaction, then it should be possible to
understand their effect in terms of the local equations of homogeneous
and isotropic cosmology \cite{Brandenberger:2004ix}. Our approach
is to suppose that a second order perturbative computation has been
performed; the homogeneous component then represents the backreaction
on the bare cosmological quantities.

In the next section, we deduce the general form of homogeneous scalar
metric perturbations in an FLRW universe. We introduce the uniform
curvature and conformal gauges in section \ref{sec:Gauge-invariance}
and construct the gauge invariant metric function that will be crucial
in our discussion of backreaction. In section \ref{sec:A-comment-on},
we point out the weakness of the longitudinal gauge in the homogeneous
limit and show that the uniform curvature gauge encodes the physical
metric degree of freedom at all wavelengths. We also write down the
perturbed Einstein equations for hydrodynamical matter in the uniform
curvature gauge. In section \ref{sec:Cosmological-backreaction},
treating homogeneous perturbations as arising at second order, we
argue that backreaction is quantified by transforming to the conformal
gauge and that, consequently, time and the scale factor are rescaled.
We establish a rule of thumb for backreaction to hasten or delay the
expansion of the universe, and illustrate what this implies for inflation
and late-time acceleration.

\section{Homogeneous perturbations\label{sec:Homogeneous-perturbations}}

In the standard theory of cosmological perturbations \cite{Lifshitz:1963ps,Bardeen:1980kt,Kodama:1985bj,Mukhanov:1990me},
all quantities are referred to the unperturbed manifold. Therefore,
homogeneous perturbations are those with no dependence on the spatial
coordinates of the background FLRW spacetime. One also assumes, usually
implicitly, that the spatial average of first order perturbations
vanishes. By this assumption, there can be no homogeneous perturbations
at first order. For reasons that will become clear, we consider more
general perturbations for which this assumption is relaxed. We will
require only that the perturbations are bounded in space. For simplicity,
we assume a flat model of the universe and use $\eta$ to denote conformal
time.

Homogeneous metric perturbations take the general form\begin{equation}
\delta g_{\mu\nu}=a^{2}(\eta)\left[\begin{array}{rr}
2\phi(\eta) & \Sigma_{i}(\eta)\\
\Sigma_{j}(\eta) & -H_{ij}(\eta)\end{array}\right],\end{equation}
where $a(\eta)$ is the background scale factor. These can be decomposed
into the usual scalar, vector and tensor pieces with respect to the
spatial metric $\delta_{ij}$ as follows \cite{Mukhanov:1990me}:\begin{eqnarray*}
\Sigma_{i} & = & S_{i}-B_{,i}\,,\\
H_{ij} & = & -2\psi\,\delta_{ij}+2E_{,ij}+F_{i,j}+F_{j,i}+h_{ij},\end{eqnarray*}
where $S^{i}{}_{,i}=F^{i}{}_{,i}=h^{ij}{}_{,i}=h^{i}{}_{i}=0.$ The
actual decomposition follows from the equations:\begin{eqnarray*}
\nabla^{2}\psi & = & 0,\\
\nabla^{2}E & = & 3\psi+\half H^{i}{}_{i},\\
\nabla^{2}B & = & 0,\\
\nabla^{2}F_{j} & = & -4\psi_{,j},\end{eqnarray*}
where $\nabla^{2}$ is the Laplacian operator on the 3-metric $\delta_{ij}.$
The appropriate boundary conditions for these equations are that the
quantities remain bounded. It then follows that all are simply functions
of time and that the right hand sides must vanish. In particular,
we have $\psi=-\myfrac16H^{i}{}_{i},$ and so $S_{i}=\Sigma_{i}$
and $h_{ij}=H_{ij}-\myfrac13H^{k}{}_{k}\delta_{ij}.$ We conclude
that the scalar part of the metric perturbations has the form\begin{equation}
\delta g_{\mu\nu}^{(S)}=a^{2}\left[\begin{array}{cc}
2\phi & 0\\
0 & 2\psi\,\delta_{ij}\end{array}\right].\end{equation}

\section{Gauge invariance\label{sec:Gauge-invariance}}

From now on we will concentrate on scalar perturbations. As usual
we will want to choose a gauge that exploits gauge-invariant quantities.
It turns out that the longitudinal gauge is not a useful gauge here;
instead we will consider the uniform curvature gauge and the conformal
gauge.

Because $B$ and $E$ have no spatial dependence, they are effectively
zero as far as the metric is concerned. But this does not mean we
are in the longitudinal gauge. The coordinate transformation $x^{\mu}\rightarrow\overline{x}^{\mu}=x^{\mu}+\delta x^{\mu}$
that affects the scalar perturbations has $\delta x^{\mu}=(\xi^{0},\xi^{,i})$
and leads to \cite{Mukhanov:1990me}\begin{eqnarray*}
\phi & \rightarrow & \overline{\phi}=\phi-\frac{1}{a}(a\xi^{0})',\\
\psi & \rightarrow & \overline{\psi}=\psi+\frac{a'}{a}\xi^{0},\\
B & \rightarrow & \overline{B}=B+\xi^{0}-\xi',\\
E & \rightarrow & \overline{E}=E-\xi,\end{eqnarray*}
where $'$ denotes a derivative with respect to $\eta.$ We can choose
the longitudinal gauge $(\overline{E}=\overline{B}=0)$ in the usual
way: $\xi=E,\,\xi^{0}=E'-B,$ but this does not simplify the form
of the metric at all. In fact, as long as $\xi^{0}$ and $\xi$ are
functions of $\eta,$ $B$ and $E$ remain effectively zero. Since
the choice of $\xi$ has no bearing on $\phi$ and $\psi,$ we conclude
that a residual degree of freedom, namely $\xi^{0},$ exists for homogeneous
perturbations in the longitudinal gauge. Therefore, there will be
a gauge artifact in our solutions if we simply let $k\rightarrow0$
in the standard treatment of cosmological perturbations. We will return
to this point shortly.

A gauge-invariant quantity can be constructed solely out of $\phi$
and $\psi.$ The most convenient choice is\begin{equation}
\Omega=\phi+\frac{1}{a}\left(\frac{a\psi}{\mathcal{H}}\right)',\end{equation}
where $\mathcal{H}=a'/a.$ Furthermore, since perturbations of scalars
transform as $\delta q\rightarrow\overline{\delta q}=\delta q-q_{0}'\xi^{0},$
\begin{equation}
\delta q^{(gi)}=\delta q+\frac{q_{0}'\psi}{\mathcal{H}}\end{equation}
will also be gauge-invariant. It is straightforward to show that in
terms of the gauge-invariant quantities natural to the longitudinal
gauge $\delta q^{(gi)}=\delta q^{(LG)}+q_{0}'\Psi/\mathcal{H}.$

\subsection{Uniform curvature gauge}

Choosing $\xi^{0}=-\psi/\mathcal{H},$ completely fixes the gauge
and gives $\overline{\psi}=0$ and $\overline{\phi}=\Omega.$ Therefore,
only the $g_{00}$ component of the metric is perturbed and all scalar
perturbations are automatically gauge-invariant. Because $\overline{\psi}=0$
this gauge is called the uniform curvature gauge or the spatially
flat gauge.

A perfect fluid is characterized by its density $\rho,$ pressure
$p$ and 3-velocity $u^{i}.$ In the uniform curvature gauge and to
first order in homogeneous perturbations, the Einstein equations become\begin{eqnarray}
-3\mathcal{H}^{2}\Omega & = & 4\pi Ga^{2}\delta\rho^{(gi)},\label{hom1}\\
0 & = & \delta u_{i}^{(gi)},\label{hom2}\\
(2\mathcal{H}'+\mathcal{H}^{2})\Omega+\mathcal{H}\Omega' & = & 4\pi Ga^{2}\delta p^{(gi)},\label{hom3}\end{eqnarray}
where $\delta u_{i}^{(gi)}=\delta u_{i}+a\psi_{,i}/\mathcal{H}=\delta u_{i}^{(LG)}+a\Psi_{,i}/\mathcal{H}.$
Making use of the background equations $3\mathcal{H}^{2}=8\pi Ga^{2}\rho_{0},\,\mathcal{H}'-\mathcal{H}^{2}=-4\pi Ga^{2}(\rho_{0}+p_{0})$
and $\rho_{0}'+3\mathcal{H}(\rho_{0}+p_{0})=0,$ we can rewrite the
first and third equations as\begin{eqnarray}
\Omega & = & -\half\frac{\delta\rho^{(gi)}}{\rho_{0}},\\
\delta\rho^{(gi)\prime} & = & -3\mathcal{H}(\delta\rho^{(gi)}+\delta p^{(gi)}).\end{eqnarray}
 As an example, we consider the case of adiabatic perturbations: $\delta p^{(gi)}=(p_{0}'/\rho_{0}')\,\delta\rho^{(gi)}$.
Then\begin{equation}
\Omega=-\myfrac32\zeta\left(1+\frac{p_{0}}{\rho_{0}}\right),\end{equation}
where \begin{equation}
\zeta\equiv-\frac{\mathcal{H}\delta\rho^{(gi)}}{\rho_{0}'}=-\Psi-\frac{\mathcal{H}\delta\rho^{(LG)}}{\rho_{0}'}\end{equation}
 is the well known invariant from cosmological perturbation theory.
Here $\zeta=\mbox{const.}$

\subsection{Conformal gauge}

If we choose $\xi^{0}=\int\! d\eta\,(\phi+\psi)$, then $\overline{\phi}=-\overline{\psi}$
and the metric is returned to a conformal FLRW metric. Although the
conformal gauge is not a good gauge---$\xi^{0}$ is only known up
to an additive constant---it suggests a way to quantify the backreaction
of perturbations on the background scale factor. We will develop the
idea fully in section \ref{sec:Cosmological-backreaction}.

\section{A comment on $k=0$ vs. $k\rightarrow0$\label{sec:A-comment-on}}

The standard treatment of (inhomogeneous) scalar perturbations shows
there is essentially one metric scalar degree of freedom: there are
two gauge-invariant scalars with one constraint. To be precise, there
is a combination of the two scalars, call it $D,$ which satisfies
a set of purely spatial differential equations. The $k\rightarrow0$
prescription is to find $D$ and then set the spatial derivatives
to zero in the remaining equations. On the other hand, choosing $k=0$
means dropping all spatial derivatives from the start and so there
is no hope of subsequently finding $D.$ The weakness of the longitudinal
gauge is that it requires knowledge of $D$ to be able to fully specify
homogeneous perturbations. This is exactly the gauge artifact mentioned
previously. Fortunately, as we now show, the uniform curvature gauge
is independent of $D$ in the homogeneous limit. In other words, $k=0$
and $k\rightarrow0$ are the same in the uniform curvature gauge.
We will illustrate this fact by again considering the case of a perfect
fluid. 

The uniform curvature gauge can be extended to general perturbations
\cite{Lukash:1980iv,Bardeen:1980kt,Lyth:1984gv,Kodama:1985bj} by
choosing $\xi=E$ so that $\overline{E}=0$ in addition to $\overline{\psi}=0.$
In this way, the scalar metric perturbations are confined solely to
the time-time and time-space components of the metric. We also introduce
the gauge-invariant combination\begin{equation}
\beta=B-E'-\frac{\psi}{\mathcal{H}}.\end{equation}
To facilitate comparison with the longitudinal gauge, in which the
scalar perturbations $\phi$ and $\psi$ become the usual gauge-invariant
quantities $\Phi$ and $\Psi,$ we note that\begin{equation}
\left.\begin{array}{l}
\Phi=\Omega+\beta'+\mathcal{H}\beta\\
\Psi=-\mathcal{H}\beta\end{array}\right\} \Longleftrightarrow\left\{ \begin{array}{l}
\Omega=\Phi+\Psi+\left(\frac{\Psi}{\mathcal{H}}\right)'\\
\beta=-\frac{\Psi}{\mathcal{H}}\end{array}\right..\end{equation}

The perturbed equations for a perfect fluid in the uniform curvature
gauge are quite compact:\begin{eqnarray}
-3\mathcal{H}^{2}\Omega-\mathcal{H}\nabla^{2}\beta & = & 4\pi Ga^{2}\delta\rho^{(gi)},\\
\mathcal{H}\Omega_{,i} & = & 4\pi Ga(\rho_{0}+p_{0})\delta u_{i}^{(gi)},\\
\left[(2\mathcal{H}'+\mathcal{H}^{2})\Omega+\mathcal{H}\Omega'\right.\nonumber \\
\left.+\half\nabla^{2}D\right]\delta_{j}^{i}-\half D^{,i}{}_{,j} & = & 4\pi Ga^{2}\delta p^{(gi)}\delta_{j}^{i},\label{Deqn}\end{eqnarray}
where $D=\Omega+\beta'+2\mathcal{H}\beta=\Phi-\Psi.$ The off-diagonal
terms and the differences between the diagonal terms of (\ref{Deqn})
imply that, for $D$ bounded, $D=D(\eta).$ If the spatial average
of $D$ is to vanish then we must have $D=0.$ However, in the homogeneous
limit of the uniform curvature gauge, we do not need to know $D.$
This is because the term in $\beta$ also drops out; we return to
eqns. (\ref{hom1})-(\ref{hom3}) and $\Omega$ is the one metric
degree of freedom required. In contrast, in the longitudinal gauge,
terms in both $\Phi$ and $\Psi$ remain if $k=0.$ We conclude that
$\Omega$ is the true physical degree of freedom describing long-wavelength
fluctuations and that the discussion of homogeneous perturbations
is best had in the uniform curvature gauge. The uniform curvature
gauge can also be extended to second order \cite{Lyth:2003im,Malik:2003mv,Bartolo:2004if,Malik:2005cy}.

Pertinent to this discussion, one of the conservation laws that can
be derived from the above equations is

\begin{equation}
\zeta'-\myfrac13\nabla^{2}\beta-\myfrac13a^{-1}\nabla^{i}\delta u_{i}^{(gi)}=-\frac{\mathcal{H}}{\rho_{0}+p_{0}}\delta p_{\mbox{\footnotesize nad}},\end{equation}
where $\delta p_{\mbox{\footnotesize nad}}=\delta p^{(gi)}-(p_{0}'/\rho_{0}')\,\delta\rho^{(gi)}$
is the non-adiabatic pressure fluctuation.

\section{Cosmological backreaction\label{sec:Cosmological-backreaction}}

All cosmological observations are carried out assuming an underlying
FLRW metric. The logic of backreaction is that, while perturbation
theory remains valid, higher order quantities affect the evolution
of lower order quantities. As emphasized earlier, this implies, for
example, that $a(\eta)$ is the bare scale factor; the actual observed
scale factor is found after backreaction has been taken into account.
We are looking for the homogeneous effect of second order terms on
background quantities.

In this light, the homogeneous perturbations in section \ref{sec:Homogeneous-perturbations}
should be thought of as second order perturbations; first order quantities
vanish in the homogeneous limit. As an example, if $q=q_{0}+q_{1}+q_{2}+\ldots$
is a perturbative expansion then, generically, we would expect the
homogeneous perturbation\[
\delta q\sim\langle q_{1}^{2}\rangle.\]
Note that this does not vanish, in contrast to the term $\langle q_{0}q_{2}\rangle=q_{0}\langle q_{2}\rangle=0.$
Furthermore, the observed scale factor can then be found by converting
to the conformal gauge. Unfortunately, because the conformal gauge
is not fixed, this procedure is not unique. We claim that it is most
consistent to perform computations in the uniform curvature gauge
and from there convert to the conformal gauge ($\phi\rightarrow\Omega,\,\psi\rightarrow0$).

To be specific, $\xi^{0}=\int\! d\eta\,\Omega$ and all background
quantities should be written in terms of the rescaled conformal time
$\overline{\eta}$ via\begin{equation}
\eta=\overline{\eta}-\!\int\! d\eta\,\Omega(\eta)\simeq\overline{\eta}-\!\int\! d\overline{\eta}\,\Omega(\overline{\eta}).\label{newtime}\end{equation}
Such rescaling of the time is a common technique in nonlinear perturbation
theory and is similar to the approaches found in \cite{Brandenberger:2004ix,Nambu:2005ne,Bartolo:2005fp}.
The fact that $\xi^{0}$ is only specified up to a constant is not
a serious drawback; it will be fixed once the initial time is chosen,
e.g. when $a(\eta_{0})=1.$ Apart from the rescaling of the time,
this integration constant does not affect the scale factor nor the
Hubble and deceleration parameters:

\begin{eqnarray}
\overline{a}(\overline{\eta}) & = & a(\eta)=a(\overline{\eta}-\xi^{0}),\\
\overline{\mathcal{H}}(\overline{\eta}) & = & \mathcal{H}(\eta)\left(1-\Omega\right),\\
\overline{q}(\overline{\eta}) & = & q(\eta)+\frac{\Omega'}{\mathcal{H}(\eta)}.\end{eqnarray}

Equation (\ref{newtime}) suggests a useful rule of thumb: cosmological
evolution will be delayed (hastened) when $\int\! d\eta\,\Omega(\eta)$
is positive (negative).

To add some physical content to this statement, let us suppose second
order effects can be treated like a perfect fluid component. Such
a situation was shown to arise in inflationary backreaction \cite{Abramo:1997hu,Geshnizjani:2002wp,Geshnizjani:2003cn}:
the second order terms contribute to an effective energy-stress-momentum
tensor that is added to the right hand side of the unperturbed Einstein
equations. In this particular case, $\delta p^{(gi)}\simeq-\delta\rho^{(gi)}\simeq\mbox{const.}>0.$
Since $\rho_{0}$ is also roughly constant but positive during inflation,
we have from eq. (\ref{hom1}),\begin{equation}
\int\! d\eta\,\Omega\simeq-\half\frac{\delta\rho^{(gi)}}{\rho_{0}}\eta>0.\end{equation}
We conclude, as in \cite{Abramo:1997hu}, that backreaction works
to slow inflationary expansion.

A question currently of some interest is whether backreaction can
lead to late-time acceleration of the universe. Since a stage of matter
domination has $q=\half$ and de Sitter expansion $q=-1,$ we can
already see that such a transition would necessarily require the breakdown
of perturbation theory. Nevertheless, our prescription can indicate
the tendency of backreaction to accelerate the expansion of the universe.
Taking $p_{0}=0$ for dust so that the background scale factor is
$a\sim\eta^{2},$ we can say for certain that we require negative
non-adiabatic pressure fluctuations (see also \cite{Kolb:2004jg,Nambu:2005ne}):\begin{equation}
\overline{q}\simeq\half+\frac{\Omega'}{\mathcal{H}}=\half+\myfrac32\frac{\delta p_{\mbox{\footnotesize nad}}}{\rho_{0}}.\end{equation}
This should also be combined with $\delta\rho^{(gi)}>0$ so that $\int\! d\eta\,\Omega<0$
is guaranteed.

\section{Conclusions}

Since first order cosmological fluctuations are required to vanish
under spatial averaging, homogeneous perturbations can only arise
at second order. They then represent corrections to the background
quantities. We formulated a gauge-invariant description of homogeneous
perturbations in order to quantify this backreaction. The uniform
curvature gauge is the most convenient gauge to exploit this gauge
invariance and we found that the gauge-variant metric function $\Omega$
is the true physical degree of freedom in the homogeneous limit. The
final step is to transform to the conformal gauge to deduce the actual
observed background quantities. We saw that this amounted to a rescaling
of the time coordinate, an outcome which is well known in nonlinear
perturbation theory. The value of the rule of thumb presented here
is that once second order perturbations have been computed---admittedly
a non-trivial task---one can quickly see what effect they have on
the evolution of the universe. The gauge invariant quantity $\Omega$
also clarifies issues surrounding $k=0$ vs. $k\rightarrow0$ and
suggests new metric ansätze to study fully nonlinear perturbations. 

\begin{acknowledgments}
It is a pleasure to thank L. R. W. Abramo, T. Buchert and S. Winitzki
for their insightful comments. MP was supported by Sonderforschungsbereich
SFB 375 für Astro--Teilchenphysik der Deutschen Forschungsgemeinschaft.

\bibliographystyle{plain}
\bibliography{/amnt/home/parry/.lyx/citations}

\begin{thebibliography}{56}
\expandafter\ifx\csname natexlab\endcsname\relax\def\natexlab#1{#1}\fi
\expandafter\ifx\csname bibnamefont\endcsname\relax
  \def\bibnamefont#1{#1}\fi
\expandafter\ifx\csname bibfnamefont\endcsname\relax
  \def\bibfnamefont#1{#1}\fi
\expandafter\ifx\csname citenamefont\endcsname\relax
  \def\citenamefont#1{#1}\fi
\expandafter\ifx\csname url\endcsname\relax
  \def\url#1{\texttt{#1}}\fi
\expandafter\ifx\csname urlprefix\endcsname\relax\def\urlprefix{URL }\fi
\providecommand{\bibinfo}[2]{#2}
\providecommand{\eprint}[2][]{\url{#2}}

\bibitem[{\citenamefont{Spergel et~al.}(2006)}]{Spergel:2006hy}
\bibinfo{author}{\bibfnamefont{D.~N.} \bibnamefont{Spergel}}
  \bibnamefont{et~al.} (\bibinfo{year}{2006}), \eprint{astro-ph/0603449}.

\bibitem[{\citenamefont{Acquaviva et~al.}(2003)\citenamefont{Acquaviva,
  Bartolo, Matarrese, and Riotto}}]{Acquaviva:2002ud}
\bibinfo{author}{\bibfnamefont{V.}~\bibnamefont{Acquaviva}},
  \bibinfo{author}{\bibfnamefont{N.}~\bibnamefont{Bartolo}},
  \bibinfo{author}{\bibfnamefont{S.}~\bibnamefont{Matarrese}},
  \bibnamefont{and} \bibinfo{author}{\bibfnamefont{A.}~\bibnamefont{Riotto}},
  \bibinfo{journal}{Nucl. Phys.} \textbf{\bibinfo{volume}{B667}},
  \bibinfo{pages}{119} (\bibinfo{year}{2003}), \eprint{astro-ph/0209156}.

\bibitem[{\citenamefont{Bartolo et~al.}(2004)\citenamefont{Bartolo, Komatsu,
  Matarrese, and Riotto}}]{Bartolo:2004if}
\bibinfo{author}{\bibfnamefont{N.}~\bibnamefont{Bartolo}},
  \bibinfo{author}{\bibfnamefont{E.}~\bibnamefont{Komatsu}},
  \bibinfo{author}{\bibfnamefont{S.}~\bibnamefont{Matarrese}},
  \bibnamefont{and} \bibinfo{author}{\bibfnamefont{A.}~\bibnamefont{Riotto}},
  \bibinfo{journal}{Phys. Rept.} \textbf{\bibinfo{volume}{402}},
  \bibinfo{pages}{103} (\bibinfo{year}{2004}), \eprint{astro-ph/0406398}.

\bibitem[{\citenamefont{Martineau and
  Brandenberger}(2005{\natexlab{a}})}]{Martineau:2005aa}
\bibinfo{author}{\bibfnamefont{P.}~\bibnamefont{Martineau}} \bibnamefont{and}
  \bibinfo{author}{\bibfnamefont{R.~H.} \bibnamefont{Brandenberger}},
  \bibinfo{journal}{Phys. Rev.} \textbf{\bibinfo{volume}{D72}},
  \bibinfo{pages}{023507} (\bibinfo{year}{2005}{\natexlab{a}}),
  \eprint{astro-ph/0505236}.

\bibitem[{\citenamefont{Bartolo et~al.}(2005)\citenamefont{Bartolo, Matarrese,
  and Riotto}}]{Bartolo:2005fp}
\bibinfo{author}{\bibfnamefont{N.}~\bibnamefont{Bartolo}},
  \bibinfo{author}{\bibfnamefont{S.}~\bibnamefont{Matarrese}},
  \bibnamefont{and} \bibinfo{author}{\bibfnamefont{A.}~\bibnamefont{Riotto}},
  \bibinfo{journal}{JCAP} \textbf{\bibinfo{volume}{0508}}, \bibinfo{pages}{010}
  (\bibinfo{year}{2005}), \eprint{astro-ph/0506410}.

\bibitem[{\citenamefont{Finelli et~al.}(2006)\citenamefont{Finelli, Marozzi,
  Vacca, and Venturi}}]{Finelli:2006wk}
\bibinfo{author}{\bibfnamefont{F.}~\bibnamefont{Finelli}},
  \bibinfo{author}{\bibfnamefont{G.}~\bibnamefont{Marozzi}},
  \bibinfo{author}{\bibfnamefont{G.~P.} \bibnamefont{Vacca}}, \bibnamefont{and}
  \bibinfo{author}{\bibfnamefont{G.}~\bibnamefont{Venturi}}
  (\bibinfo{year}{2006}), \eprint{gr-qc/0604081}.

\bibitem[{\citenamefont{Futamase}(1988)}]{Futamase:1988aa}
\bibinfo{author}{\bibfnamefont{T.}~\bibnamefont{Futamase}},
  \bibinfo{journal}{Phys. Rev. Lett.} \textbf{\bibinfo{volume}{61}},
  \bibinfo{pages}{2175} (\bibinfo{year}{1988}).

\bibitem[{\citenamefont{Futamase}(1989)}]{Futamase:1989aa}
\bibinfo{author}{\bibfnamefont{T.}~\bibnamefont{Futamase}},
  \bibinfo{journal}{Mon. Not. R. astr. Soc.} \textbf{\bibinfo{volume}{237}},
  \bibinfo{pages}{187} (\bibinfo{year}{1989}).

\bibitem[{\citenamefont{Bildhauer and Futamase}(1991)}]{Bildhauer:1991aa}
\bibinfo{author}{\bibfnamefont{S.}~\bibnamefont{Bildhauer}} \bibnamefont{and}
  \bibinfo{author}{\bibfnamefont{T.}~\bibnamefont{Futamase}},
  \bibinfo{journal}{Gen. Rel. Grav.} \textbf{\bibinfo{volume}{23}},
  \bibinfo{pages}{1251} (\bibinfo{year}{1991}).

\bibitem[{\citenamefont{Zalaletdinov}(1992)}]{Zalaletdinov:1992cg}
\bibinfo{author}{\bibfnamefont{R.~M.} \bibnamefont{Zalaletdinov}},
  \bibinfo{journal}{Gen. Rel. Grav.} \textbf{\bibinfo{volume}{24}},
  \bibinfo{pages}{1015} (\bibinfo{year}{1992}).

\bibitem[{\citenamefont{Buchert and Ehlers}(1997)}]{Buchert:1995fz}
\bibinfo{author}{\bibfnamefont{T.}~\bibnamefont{Buchert}} \bibnamefont{and}
  \bibinfo{author}{\bibfnamefont{J.}~\bibnamefont{Ehlers}},
  \bibinfo{journal}{Astron. Astrophys.} \textbf{\bibinfo{volume}{320}},
  \bibinfo{pages}{1} (\bibinfo{year}{1997}), \eprint{astro-ph/9510056}.

\bibitem[{\citenamefont{Russ et~al.}(1997)\citenamefont{Russ, Soffel, Kasai,
  and B{\" o}rner}}]{Russ:1996km}
\bibinfo{author}{\bibfnamefont{H.}~\bibnamefont{Russ}},
  \bibinfo{author}{\bibfnamefont{M.~H.} \bibnamefont{Soffel}},
  \bibinfo{author}{\bibfnamefont{M.}~\bibnamefont{Kasai}}, \bibnamefont{and}
  \bibinfo{author}{\bibfnamefont{G.}~\bibnamefont{B{\" o}rner}},
  \bibinfo{journal}{Phys. Rev.} \textbf{\bibinfo{volume}{D56}},
  \bibinfo{pages}{2044} (\bibinfo{year}{1997}), \eprint{astro-ph/9612218}.

\bibitem[{\citenamefont{Zalaletdinov}(1997)}]{Zalaletdinov:1996aj}
\bibinfo{author}{\bibfnamefont{R.~M.} \bibnamefont{Zalaletdinov}},
  \bibinfo{journal}{Bull. Astron. Soc. India} \textbf{\bibinfo{volume}{25}},
  \bibinfo{pages}{401} (\bibinfo{year}{1997}), \eprint{gr-qc/9703016}.

\bibitem[{\citenamefont{Buchert}(2000)}]{Buchert:1999er}
\bibinfo{author}{\bibfnamefont{T.}~\bibnamefont{Buchert}},
  \bibinfo{journal}{Gen. Rel. Grav.} \textbf{\bibinfo{volume}{32}},
  \bibinfo{pages}{105} (\bibinfo{year}{2000}), \eprint{gr-qc/9906015}.

\bibitem[{\citenamefont{Buchert}(2001)}]{Buchert:2001sa}
\bibinfo{author}{\bibfnamefont{T.}~\bibnamefont{Buchert}},
  \bibinfo{journal}{Gen. Rel. Grav.} \textbf{\bibinfo{volume}{33}},
  \bibinfo{pages}{1381} (\bibinfo{year}{2001}), \eprint{gr-qc/0102049}.

\bibitem[{\citenamefont{Buchert and Carfora}(2002)}]{Buchert:2002ht}
\bibinfo{author}{\bibfnamefont{T.}~\bibnamefont{Buchert}} \bibnamefont{and}
  \bibinfo{author}{\bibfnamefont{M.}~\bibnamefont{Carfora}},
  \bibinfo{journal}{Class. Quant. Grav.} \textbf{\bibinfo{volume}{19}},
  \bibinfo{pages}{6109} (\bibinfo{year}{2002}), \eprint{gr-qc/0210037}.

\bibitem[{\citenamefont{Buchert and Carfora}(2003)}]{Buchert:2002ij}
\bibinfo{author}{\bibfnamefont{T.}~\bibnamefont{Buchert}} \bibnamefont{and}
  \bibinfo{author}{\bibfnamefont{M.}~\bibnamefont{Carfora}},
  \bibinfo{journal}{Phys. Rev. Lett.} \textbf{\bibinfo{volume}{90}},
  \bibinfo{pages}{031101} (\bibinfo{year}{2003}), \eprint{gr-qc/0210045}.

\bibitem[{\citenamefont{Ellis and Buchert}(2005)}]{Ellis:2005uz}
\bibinfo{author}{\bibfnamefont{G.~F.~R.} \bibnamefont{Ellis}} \bibnamefont{and}
  \bibinfo{author}{\bibfnamefont{T.}~\bibnamefont{Buchert}},
  \bibinfo{journal}{Phys. Lett.} \textbf{\bibinfo{volume}{A347}},
  \bibinfo{pages}{38} (\bibinfo{year}{2005}), \eprint{gr-qc/0506106}.

\bibitem[{\citenamefont{Mars and Zalaletdinov}(1997)}]{Mars:1997jy}
\bibinfo{author}{\bibfnamefont{M.}~\bibnamefont{Mars}} \bibnamefont{and}
  \bibinfo{author}{\bibfnamefont{R.~M.} \bibnamefont{Zalaletdinov}},
  \bibinfo{journal}{J. Math. Phys.} \textbf{\bibinfo{volume}{38}},
  \bibinfo{pages}{4741} (\bibinfo{year}{1997}), \eprint{dg-ga/9703002}.

\bibitem[{\citenamefont{Mukhanov et~al.}(1997)\citenamefont{Mukhanov, Abramo,
  and Brandenberger}}]{Mukhanov:1996ak}
\bibinfo{author}{\bibfnamefont{V.~F.} \bibnamefont{Mukhanov}},
  \bibinfo{author}{\bibfnamefont{L.~R.~W.} \bibnamefont{Abramo}},
  \bibnamefont{and} \bibinfo{author}{\bibfnamefont{R.~H.}
  \bibnamefont{Brandenberger}}, \bibinfo{journal}{Phys. Rev. Lett.}
  \textbf{\bibinfo{volume}{78}}, \bibinfo{pages}{1624} (\bibinfo{year}{1997}),
  \eprint{gr-qc/9609026}.

\bibitem[{\citenamefont{Bruni et~al.}(1997)\citenamefont{Bruni, Matarrese,
  Mollerach, and Sonego}}]{Bruni:1996im}
\bibinfo{author}{\bibfnamefont{M.}~\bibnamefont{Bruni}},
  \bibinfo{author}{\bibfnamefont{S.}~\bibnamefont{Matarrese}},
  \bibinfo{author}{\bibfnamefont{S.}~\bibnamefont{Mollerach}},
  \bibnamefont{and} \bibinfo{author}{\bibfnamefont{S.}~\bibnamefont{Sonego}},
  \bibinfo{journal}{Class. Quant. Grav.} \textbf{\bibinfo{volume}{14}},
  \bibinfo{pages}{2585} (\bibinfo{year}{1997}), \eprint{gr-qc/9609040}.

\bibitem[{\citenamefont{Abramo et~al.}(1997)\citenamefont{Abramo,
  Brandenberger, and Mukhanov}}]{Abramo:1997hu}
\bibinfo{author}{\bibfnamefont{L.~R.~W.} \bibnamefont{Abramo}},
  \bibinfo{author}{\bibfnamefont{R.~H.} \bibnamefont{Brandenberger}},
  \bibnamefont{and} \bibinfo{author}{\bibfnamefont{V.~F.}
  \bibnamefont{Mukhanov}}, \bibinfo{journal}{Phys. Rev.}
  \textbf{\bibinfo{volume}{D56}}, \bibinfo{pages}{3248} (\bibinfo{year}{1997}),
  \eprint{gr-qc/9704037}.

\bibitem[{\citenamefont{Rigopoulos}(2004)}]{Rigopoulos:2002mc}
\bibinfo{author}{\bibfnamefont{G.}~\bibnamefont{Rigopoulos}},
  \bibinfo{journal}{Class. Quant. Grav.} \textbf{\bibinfo{volume}{21}},
  \bibinfo{pages}{1737} (\bibinfo{year}{2004}), \eprint{astro-ph/0212141}.

\bibitem[{\citenamefont{Lyth and Wands}(2003)}]{Lyth:2003im}
\bibinfo{author}{\bibfnamefont{D.~H.} \bibnamefont{Lyth}} \bibnamefont{and}
  \bibinfo{author}{\bibfnamefont{D.}~\bibnamefont{Wands}},
  \bibinfo{journal}{Phys. Rev.} \textbf{\bibinfo{volume}{D68}},
  \bibinfo{pages}{103515} (\bibinfo{year}{2003}), \eprint{astro-ph/0306498}.

\bibitem[{\citenamefont{Noh and Hwang}(2004)}]{Noh:2004bc}
\bibinfo{author}{\bibfnamefont{H.}~\bibnamefont{Noh}} \bibnamefont{and}
  \bibinfo{author}{\bibfnamefont{J.-c.} \bibnamefont{Hwang}},
  \bibinfo{journal}{Phys. Rev.} \textbf{\bibinfo{volume}{D69}},
  \bibinfo{pages}{104011} (\bibinfo{year}{2004}).

\bibitem[{\citenamefont{Malik and Wands}(2004)}]{Malik:2003mv}
\bibinfo{author}{\bibfnamefont{K.~A.} \bibnamefont{Malik}} \bibnamefont{and}
  \bibinfo{author}{\bibfnamefont{D.}~\bibnamefont{Wands}},
  \bibinfo{journal}{Class. Quant. Grav.} \textbf{\bibinfo{volume}{21}},
  \bibinfo{pages}{L65} (\bibinfo{year}{2004}), \eprint{astro-ph/0307055}.

\bibitem[{\citenamefont{Vernizzi}(2005)}]{Vernizzi:2004nc}
\bibinfo{author}{\bibfnamefont{F.}~\bibnamefont{Vernizzi}},
  \bibinfo{journal}{Phys. Rev.} \textbf{\bibinfo{volume}{D71}},
  \bibinfo{pages}{061301} (\bibinfo{year}{2005}), \eprint{astro-ph/0411463}.

\bibitem[{\citenamefont{Malik}(2005)}]{Malik:2005cy}
\bibinfo{author}{\bibfnamefont{K.~A.} \bibnamefont{Malik}},
  \bibinfo{journal}{JCAP} \textbf{\bibinfo{volume}{0511}}, \bibinfo{pages}{005}
  (\bibinfo{year}{2005}), \eprint{astro-ph/0506532}.

\bibitem[{\citenamefont{Brandenberger}(1999)}]{Brandenberger:1999su}
\bibinfo{author}{\bibfnamefont{R.~H.} \bibnamefont{Brandenberger}}
  (\bibinfo{year}{1999}), \eprint{hep-th/0004016}.

\bibitem[{\citenamefont{Brandenberger}(2002)}]{Brandenberger:2002sk}
\bibinfo{author}{\bibfnamefont{R.~H.} \bibnamefont{Brandenberger}}
  (\bibinfo{year}{2002}), \eprint{hep-th/0210165}.

\bibitem[{\citenamefont{Unruh}(1998)}]{Unruh:1998ic}
\bibinfo{author}{\bibfnamefont{W.}~\bibnamefont{Unruh}} (\bibinfo{year}{1998}),
  \eprint{astro-ph/9802323}.

\bibitem[{\citenamefont{Abramo and Woodard}(2002)}]{Abramo:2001dc}
\bibinfo{author}{\bibfnamefont{L.~R.} \bibnamefont{Abramo}} \bibnamefont{and}
  \bibinfo{author}{\bibfnamefont{R.~P.} \bibnamefont{Woodard}},
  \bibinfo{journal}{Phys. Rev.} \textbf{\bibinfo{volume}{D65}},
  \bibinfo{pages}{063515} (\bibinfo{year}{2002}), \eprint{astro-ph/0109272}.

\bibitem[{\citenamefont{Afshordi and Brandenberger}(2001)}]{Afshordi:2000nr}
\bibinfo{author}{\bibfnamefont{N.}~\bibnamefont{Afshordi}} \bibnamefont{and}
  \bibinfo{author}{\bibfnamefont{R.~H.} \bibnamefont{Brandenberger}},
  \bibinfo{journal}{Phys. Rev.} \textbf{\bibinfo{volume}{D63}},
  \bibinfo{pages}{123505} (\bibinfo{year}{2001}), \eprint{gr-qc/0011075}.

\bibitem[{\citenamefont{Geshnizjani and
  Brandenberger}(2002)}]{Geshnizjani:2002wp}
\bibinfo{author}{\bibfnamefont{G.}~\bibnamefont{Geshnizjani}} \bibnamefont{and}
  \bibinfo{author}{\bibfnamefont{R.}~\bibnamefont{Brandenberger}},
  \bibinfo{journal}{Phys. Rev.} \textbf{\bibinfo{volume}{D66}},
  \bibinfo{pages}{123507} (\bibinfo{year}{2002}), \eprint{gr-qc/0204074}.

\bibitem[{\citenamefont{Geshnizjani and
  Brandenberger}(2005)}]{Geshnizjani:2003cn}
\bibinfo{author}{\bibfnamefont{G.}~\bibnamefont{Geshnizjani}} \bibnamefont{and}
  \bibinfo{author}{\bibfnamefont{R.}~\bibnamefont{Brandenberger}},
  \bibinfo{journal}{JCAP} \textbf{\bibinfo{volume}{0504}}, \bibinfo{pages}{006}
  (\bibinfo{year}{2005}), \eprint{hep-th/0310265}.

\bibitem[{\citenamefont{Riess et~al.}(1998)}]{Riess:1998cb}
\bibinfo{author}{\bibfnamefont{A.~G.} \bibnamefont{Riess}} \bibnamefont{et~al.}
  (\bibinfo{collaboration}{Supernova Search Team}), \bibinfo{journal}{Astron.
  J.} \textbf{\bibinfo{volume}{116}}, \bibinfo{pages}{1009}
  (\bibinfo{year}{1998}), \eprint{astro-ph/9805201}.

\bibitem[{\citenamefont{Perlmutter et~al.}(1999)}]{Perlmutter:1998np}
\bibinfo{author}{\bibfnamefont{S.}~\bibnamefont{Perlmutter}}
  \bibnamefont{et~al.} (\bibinfo{collaboration}{Supernova Cosmology Project}),
  \bibinfo{journal}{Astrophys. J.} \textbf{\bibinfo{volume}{517}},
  \bibinfo{pages}{565} (\bibinfo{year}{1999}), \eprint{astro-ph/9812133}.

\bibitem[{\citenamefont{Kolb et~al.}(2005{\natexlab{a}})\citenamefont{Kolb,
  Matarrese, Notari, and Riotto}}]{Kolb:2004am}
\bibinfo{author}{\bibfnamefont{E.~W.} \bibnamefont{Kolb}},
  \bibinfo{author}{\bibfnamefont{S.}~\bibnamefont{Matarrese}},
  \bibinfo{author}{\bibfnamefont{A.}~\bibnamefont{Notari}}, \bibnamefont{and}
  \bibinfo{author}{\bibfnamefont{A.}~\bibnamefont{Riotto}},
  \bibinfo{journal}{Phys. Rev.} \textbf{\bibinfo{volume}{D71}},
  \bibinfo{pages}{023524} (\bibinfo{year}{2005}{\natexlab{a}}),
  \eprint{hep-ph/0409038}.

\bibitem[{\citenamefont{Kolb et~al.}(2005{\natexlab{b}})\citenamefont{Kolb,
  Matarrese, Notari, and Riotto}}]{Kolb:2004jg}
\bibinfo{author}{\bibfnamefont{E.~W.} \bibnamefont{Kolb}},
  \bibinfo{author}{\bibfnamefont{S.}~\bibnamefont{Matarrese}},
  \bibinfo{author}{\bibfnamefont{A.}~\bibnamefont{Notari}}, \bibnamefont{and}
  \bibinfo{author}{\bibfnamefont{A.}~\bibnamefont{Riotto}},
  \bibinfo{journal}{Mod. Phys. Lett.} \textbf{\bibinfo{volume}{A20}},
  \bibinfo{pages}{2705} (\bibinfo{year}{2005}{\natexlab{b}}),
  \eprint{astro-ph/0410541}.

\bibitem[{\citenamefont{Flanagan}(2005)}]{Flanagan:2005dk}
\bibinfo{author}{\bibfnamefont{E.~E.} \bibnamefont{Flanagan}},
  \bibinfo{journal}{Phys. Rev.} \textbf{\bibinfo{volume}{D71}},
  \bibinfo{pages}{103521} (\bibinfo{year}{2005}), \eprint{hep-th/0503202}.

\bibitem[{\citenamefont{Geshnizjani et~al.}(2005)\citenamefont{Geshnizjani,
  Chung, and Afshordi}}]{Geshnizjani:2005ce}
\bibinfo{author}{\bibfnamefont{G.}~\bibnamefont{Geshnizjani}},
  \bibinfo{author}{\bibfnamefont{D.~J.~H.} \bibnamefont{Chung}},
  \bibnamefont{and} \bibinfo{author}{\bibfnamefont{N.}~\bibnamefont{Afshordi}},
  \bibinfo{journal}{Phys. Rev.} \textbf{\bibinfo{volume}{D72}},
  \bibinfo{pages}{023517} (\bibinfo{year}{2005}), \eprint{astro-ph/0503553}.

\bibitem[{\citenamefont{Ishibashi and Wald}(2006)}]{Ishibashi:2005sj}
\bibinfo{author}{\bibfnamefont{A.}~\bibnamefont{Ishibashi}} \bibnamefont{and}
  \bibinfo{author}{\bibfnamefont{R.~M.} \bibnamefont{Wald}},
  \bibinfo{journal}{Class. Quant. Grav.} \textbf{\bibinfo{volume}{23}},
  \bibinfo{pages}{235} (\bibinfo{year}{2006}), \eprint{gr-qc/0509108}.

\bibitem[{\citenamefont{Buchert}(2006)}]{Buchert:2005kj}
\bibinfo{author}{\bibfnamefont{T.}~\bibnamefont{Buchert}},
  \bibinfo{journal}{Class. Quant. Grav.} \textbf{\bibinfo{volume}{23}},
  \bibinfo{pages}{817} (\bibinfo{year}{2006}), \eprint{gr-qc/0509124}.

\bibitem[{\citenamefont{R{\"a}s{\"a}nen}(2006)}]{Rasanen:2005zy}
\bibinfo{author}{\bibfnamefont{S.}~\bibnamefont{R{\"a}s{\"a}nen}},
  \bibinfo{journal}{Class. Quant. Grav.} \textbf{\bibinfo{volume}{23}},
  \bibinfo{pages}{1823} (\bibinfo{year}{2006}), \eprint{astro-ph/0504005}.

\bibitem[{\citenamefont{Siegel and Fry}(2005)}]{Siegel:2005xu}
\bibinfo{author}{\bibfnamefont{E.~R.} \bibnamefont{Siegel}} \bibnamefont{and}
  \bibinfo{author}{\bibfnamefont{J.~N.} \bibnamefont{Fry}},
  \bibinfo{journal}{Astrophys. J.} \textbf{\bibinfo{volume}{628}},
  \bibinfo{pages}{L1} (\bibinfo{year}{2005}), \eprint{astro-ph/0504421}.

\bibitem[{\citenamefont{Martineau and
  Brandenberger}(2005{\natexlab{b}})}]{Martineau:2005zu}
\bibinfo{author}{\bibfnamefont{P.}~\bibnamefont{Martineau}} \bibnamefont{and}
  \bibinfo{author}{\bibfnamefont{R.}~\bibnamefont{Brandenberger}}
  (\bibinfo{year}{2005}{\natexlab{b}}), \eprint{astro-ph/0510523}.

\bibitem[{\citenamefont{Brandenberger and Lam}(2004)}]{Brandenberger:2004ix}
\bibinfo{author}{\bibfnamefont{R.~H.} \bibnamefont{Brandenberger}}
  \bibnamefont{and} \bibinfo{author}{\bibfnamefont{C.~S.} \bibnamefont{Lam}}
  (\bibinfo{year}{2004}), \eprint{hep-th/0407048}.

\bibitem[{\citenamefont{Lifshitz and Khalatnikov}(1963)}]{Lifshitz:1963ps}
\bibinfo{author}{\bibfnamefont{E.~M.} \bibnamefont{Lifshitz}} \bibnamefont{and}
  \bibinfo{author}{\bibfnamefont{I.~M.} \bibnamefont{Khalatnikov}},
  \bibinfo{journal}{Adv. Phys.} \textbf{\bibinfo{volume}{12}},
  \bibinfo{pages}{185} (\bibinfo{year}{1963}).

\bibitem[{\citenamefont{Bardeen}(1980)}]{Bardeen:1980kt}
\bibinfo{author}{\bibfnamefont{J.~M.} \bibnamefont{Bardeen}},
  \bibinfo{journal}{Phys. Rev.} \textbf{\bibinfo{volume}{D22}},
  \bibinfo{pages}{1882} (\bibinfo{year}{1980}).

\bibitem[{\citenamefont{Kodama and Sasaki}(1984)}]{Kodama:1985bj}
\bibinfo{author}{\bibfnamefont{H.}~\bibnamefont{Kodama}} \bibnamefont{and}
  \bibinfo{author}{\bibfnamefont{M.}~\bibnamefont{Sasaki}},
  \bibinfo{journal}{Prog. Theor. Phys. Suppl.} \textbf{\bibinfo{volume}{78}},
  \bibinfo{pages}{1} (\bibinfo{year}{1984}).

\bibitem[{\citenamefont{Mukhanov et~al.}(1992)\citenamefont{Mukhanov, Feldman,
  and Brandenberger}}]{Mukhanov:1990me}
\bibinfo{author}{\bibfnamefont{V.~F.} \bibnamefont{Mukhanov}},
  \bibinfo{author}{\bibfnamefont{H.~A.} \bibnamefont{Feldman}},
  \bibnamefont{and} \bibinfo{author}{\bibfnamefont{R.~H.}
  \bibnamefont{Brandenberger}}, \bibinfo{journal}{Phys. Rept.}
  \textbf{\bibinfo{volume}{215}}, \bibinfo{pages}{203} (\bibinfo{year}{1992}).

\bibitem[{\citenamefont{Lukash}(1980)}]{Lukash:1980iv}
\bibinfo{author}{\bibfnamefont{V.~N.} \bibnamefont{Lukash}},
  \bibinfo{journal}{Sov. Phys. JETP} \textbf{\bibinfo{volume}{52}},
  \bibinfo{pages}{807} (\bibinfo{year}{1980}).

\bibitem[{\citenamefont{Lyth}(1985)}]{Lyth:1984gv}
\bibinfo{author}{\bibfnamefont{D.~H.} \bibnamefont{Lyth}},
  \bibinfo{journal}{Phys. Rev.} \textbf{\bibinfo{volume}{D31}},
  \bibinfo{pages}{1792} (\bibinfo{year}{1985}).

\bibitem[{\citenamefont{Nambu}(2005)}]{Nambu:2005ne}
\bibinfo{author}{\bibfnamefont{Y.}~\bibnamefont{Nambu}},
  \bibinfo{journal}{Phys. Rev.} \textbf{\bibinfo{volume}{D71}},
  \bibinfo{pages}{084016} (\bibinfo{year}{2005}), \eprint{gr-qc/0503111}.

\bibitem[{\citenamefont{Easther and Parry}(2000)}]{Easther:1999ws}
\bibinfo{author}{\bibfnamefont{R.}~\bibnamefont{Easther}} \bibnamefont{and}
  \bibinfo{author}{\bibfnamefont{M.}~\bibnamefont{Parry}},
  \bibinfo{journal}{Phys. Rev.} \textbf{\bibinfo{volume}{D62}},
  \bibinfo{pages}{103503} (\bibinfo{year}{2000}), \eprint{hep-ph/9910441}.

\bibitem[{\citenamefont{Langlois and Vernizzi}(2005)}]{Langlois:2005qp}
\bibinfo{author}{\bibfnamefont{D.}~\bibnamefont{Langlois}} \bibnamefont{and}
  \bibinfo{author}{\bibfnamefont{F.}~\bibnamefont{Vernizzi}},
  \bibinfo{journal}{Phys. Rev.} \textbf{\bibinfo{volume}{D72}},
  \bibinfo{pages}{103501} (\bibinfo{year}{2005}), \eprint{astro-ph/0509078}.

\end{thebibliography}
\end{acknowledgments}

\end{document}